\documentclass[12pt, epsf]{article}
\usepackage{amsfonts,amssymb,amsmath,amscd, graphicx}

\title{Orbifolds in Numerical Relativity}
\author{Mihai Bondarescu
\footnote{California Institute of Technology MC 452-48, Pasadena, CA 91125, U.S.A., email 
mihai@theory.caltech.edu, web page http://theory.caltech.edu/\~{}mihai}}

\date{\today}

\begin{document}

\maketitle

\begin{abstract}

Numerical relativity has been using orbifolds for a long time, although they appear under different names in the literature. We review orbifolds previously used in simulations also discuss some that have not been used yet but are likely to be useful in the future. 

\end{abstract}


\maketitle

\section{Introduction}

An orbifold $O$ is an object that can be globally written as a coset 
\begin{equation}
O=M / G
\label{coset}
\end{equation}
where $M$ is a manifold, and $G$ is a group of its isometries, not necessarily all of them. If $G$ is trivial (contains nothing but the identity), the orbifold will be identical with the manifold ($O=M$). When $G$ is not trivial, it will map each point in $M$ to one or more points in $M$. As a result, the fundamental domain of $O$ will be significantly less than $M$. In the physics community, Orbifolds have been used extensively in String Theory \cite{Dixon:1986jc, Dixon:1985jw} and are well explained in mathematics text books \cite{BookOnOrbifolds}. 

Classical General Relativity (GR) normally studies the space we appear to live in - 4 dimensional non-compact and unorbifolded Minkowski. In practical applications, like numerical simulations, it is often convenient to consider problems with a high degree of symmetry and, in such cases, one can take advantage of these symmetries and simulate physics on an orbifold rather than full Minkowski. 

So far, orbifolds are hardly ever mentioned in the General Relativity literature.  However, orbifolds of ${\bf R}^3$ and 4 dimensional Minkowski space have been used extensively under different names in Numerical Relativity. In simulations, they help reducing the amount of computer resources (memory, time, etc) needed to evolve a given space-time, provided the initial data has enough symmetries to permit orbifolding. In Numerical Relativity, three of the most commonly used orbifolds are known as {\it bitant}, {\it quadrant} and {\it octant} grids. 

In this paper we discuss orbifolds that have been found useful in Numerical Relativity and orbifolds that have not been used yet, but could be used to get new insights into black hole physics. 

Some sets of innitial data have enough symmetry to permit orbifolding in several ways. Since each orbifold supresses a different set of unstable modes, evolving the same initial data on different orbifolds with the same fundamental domain will allow these unstable modes to be studied individually. Better understanding of unstable modes helps simulations run longer and thus helps overcome one of the most important challanges in Numerical Relativity. 

As in most of Numerical Relativity, in all but section 8 of this paper we will split the 4-dimensional space-time in three-dimensional space plus time and let the "space" evolve in time. Thus, the orbifolds of interest in this paper will be orbifolds of ${\bf R}^3$.  Of course, one can add time to these three dimensional orbifolds to obtain orbifolds of full Minkowski.

\section{Conic Orbidolds}

For any integer $n \neq 0$, a rotation by an angle $\alpha=\frac{2\pi}{n}$ in an arbitrary plane will generate a group with $n$ elements. Moding out the three dimensional space by this group or, in polar coordinates, using the identification $rot_\alpha$ given by  
\begin{equation}
(r, \theta, x_3) = (r, \theta+\alpha, x_3)
\label{conic}
\end{equation}
yields an orbifold with a singularity at $r=0$ and no other special points. Black Holes living on these orbifolds will be either centered on the singularity or away from it. Black holes living away from $r=0$ will be attracted toward $r=0$ by their own images. One can counterbalance this by adding either some charge or angular momentum in the $(r, \theta)$ plane since rotations in this plane are permitted by the topology of this orbifold.

\section{The {\it Bitant} orbifold}

The simplest non-trivial group of symmetries of ${\bf R}^3$ is

\begin{equation}
G_b=\{e,r_{x_1}\} = \langle r_{x_1} \rangle
\end{equation}
where $e$ is the identity, $r_{x_1}$ is a reflection and $\langle a \rangle $ denotes the grup generated by $a$.  Since reflecting twice gives identity, this group has only two elements. Without loss of generality, we can take the reflection to be the identification 
\begin{equation}
(x_1, x_2, x_3)=(-x_1, x_2, x_3)
\label{reflection}
\end{equation}
This orbifold will have as fundamental domain half of ${\bf R}^3$ or points $(x_1, x_2, x_3)$ such that $x_1>0$ and a  singularity on the $x_1=0$ plane. The Bitant orbifold has been used extensively for the study of space-times that exhibit the symmetry (\ref{reflection}). 

It's easy to see that the following solutions of the Einstein Equations are compatible with the topology of this orbifold.  

- flat 4 dimensional Minkowski space 

- a Schwarzschild black hole centered somewhere on the $x_1 = 0$ plane. The orbifold topology will not affect the hole in any way - not perturb the spherical symmetry, etc.

- a Kerr hole rotating in the $(x_2,x_3)$ plane or around $x_1$ and centered in the $x_1=0$ plane

- a binary black hole system composed of holes centered on the $x_1=0$ plane with spin angular momentum of the individual holes and orbital angular momenta of the binary system along $x_1$  

- two holes centered on the $x_1=0$ plane colliding head-on. 

A single Schwarzschild hole away from $x_1=0$ will never be in a stable stationary position on this orbifold. It will be attracted by its own image under the identification (\ref{reflection}).  The metric will therefore be identical with half of the spacetime of two holes colliding head-on. The end point of the evolution will be a hole centered on $x_1=0$ and emission of gravitational radiation. The orbifold topology forces the whole process to be symmetric under (\ref{reflection}).

The orbifold symmetry will not allow any solution that violates (\ref{reflection}) such as a Kerr black hole rotating in a plane containing $x_1$ and centered on the $x_1=0$ plane. 

The {\it Bitant} orbifold is also recovered from the five-dimensional solution of the Einstein Equations described in \cite{Bondarescu:2005dk} when the compact dimension is sent to zero.

{\it Bitant} grids have been used in 
\cite{Alcubierre:2002kk, Laguna:2002zc, Tiglio:2003xm} and numerous other works.

\section{Alternatives to  Bitant} 

For a rotation by angle $\alpha=\pi$ in the plane $(x_1, x_2)$, the conic orbifold obtained by moding out ${\bf R}^3$ by the 2 element group 

\begin{equation}
G_{b'}=\langle{rot_{\pi}}\rangle
\end{equation}
has the same fundamental domain as {\it Bitant}. The orbifold singularity will be the $x_1=x_2=0$ line as opposed to the $x_1=0$ plane in the case of {\it Bitant}.

 Just like {\it Bitant}, this orbifold can be used for evolving a single black hole centered at the origin, but unlike in {\it Bitant}, the black hole will not be able to wander away in the $x_2$ direction. Like {\it Bitant}, this orbifold can be used for the inspiral and collision of two equal mass holes, rotating in the $(x_1, x_2)$ plane. 

The symmetry $rot_\pi$ is currently implemented in Cactus \cite{Cactus} and is known as {\it $\pi$ symmetry}. 

Another orbifold with the same fundamental domain as {\it bitant} is obtained when one considers the symmetry 

\begin{equation} 
(x_1, x_2, x_3) = (-x_1, -x_2, -x_3)
\end{equation}
This symmetry generates a two element group and moding out ${\bf R}^3$ by this group yields an orbifold with the same fundamental domain as $\it Bitant$ but with different properties.
This orbifold will have only one singular point at the origin, thus being the least singular orbifold with half of the space as fundamental domain. 
  
This orbifold can be used for simulating the collision of equal mass opposite spin black holes. Unlike all orbifolds used so far in Numerical Relativity, this orbifold does not force the orbit of the black holes to be confined to a given plane. They can wander as they wish or as physics dictates. As the simmulation is concerned with colliding a hole with its immage, the two holes will always be on opposite sides of the origin and have opposite angular momenta. This is consistent with the conservation of total momentum.

\section {The {\it Quadrant} orbifold}

Another widely used orbifold is obtained by considering a group generated by two reflections similar to (\ref{reflection}), namely 
\begin{equation}
G_q=\langle r_{x_1}, r_{x_2} \rangle
\end{equation}
where $r_{x_1}$ is the reflection
\begin{equation}
(x_1, x_2, x_3)=(-x_1, x_2, x_3)
\end{equation}
and $r_{x_2}$ 
\begin{equation}
(x_1, x_2,x_3)=(x_1,-x_2,x_3)
\end{equation}

Here, again one can have black holes centered in the "bulk" $(x_1, x_2 \neq 0)$ or on the topological singularities  $(x_1=0, x_2=0~ or~ x_1=x_2=0)$. An uncharged non-rotating hole of mass $M$ living in the bulk centered at some point of coordinates $(x_1,x_2,x_3)$, far away from boundaries will generate a metric almost identical to Schwarzschild in its immediate neighborhood. The global metric will be that of a quadrant of the space-time generated by $4$ holes centered at $(x_1,x_2,x_3)$, $(-x_1,x_2,x_3)$, $(x_1,-x_2,x_3)$ and respectively $(-x_1,-x_2,x_3)$. The hole will not be in equilibrium. It can join one of its images and form a new hole somewhat lighter than $2M$ plus gravitational radiation. Upon collision, the space-time metric will be a quadrant of a space-time consisting of two black holes. If one allows the hole images to further coalesce, the end result will be a hole centered on the $x_1=x_2=0$ line. 

In order to save computer resources, in the work of \cite{Alcubierre:2002kk, Laguna:2002zc}, the head-on collision of 2 black holes centered on the $x_1=x_2=0$ line of the $Quadrant$ orbifold was studied.

The {\it Quadrant} orbifold does not permit rotating holes centered at $x_1=x_2=0$.

\subsection{Alternatives to {\it Quadrant}}

An alternative orbifold with the same fundamental domain as {\it Quadrant} that can be used in many of the applications where {\it Quadrant} is currently used is the conic orbifold obtained for $\alpha=\pi/2$. The corresponding group of isometries is 

\begin{equation}
G=\langle rot_{\pi/2} \rangle=\langle (x_1, x_2, x_3) = (x_2, -x_1, x_3)  \rangle
\end{equation}

Unlike {\it Quadrant}, this orbifold permits rotation in the $(x_1,x_2)$ plane and due to its different set of symmetries it will allow a different set of interesting modes to grow. 

Another alternative to {\it Quadrant} is obtained when considering the group generated by the following $\pi-$rotation  and a reflection 
\begin{equation}
(x_1,x_2,x_3)=(-x_1,-x_2,x_3)
\end{equation}
\begin{equation}
(x_1,x_2,x_3)=(x_1,x_2,-x_3)
\end{equation}

Informally, this orbifold is known as {\it  $\pi$ symmetry plus bitant} and has been used in \cite{Sperhake:2005uf, Diener:2005mg, Campanelli:2005dd} This orbifold would allow orbiting black holes or rotating binary systems as well as useful in head-on collisions to be simulated on a quadrant-size grid. 

One can combine the rotation and reflection in a different way to obtain another orbifold : 
\begin{equation}
(x_1,x_2,x_3)=(-x_1,-x_2,x_3)
\end{equation}
\begin{equation}
(x_1,x_2,x_3)=(-x_1,x_2,x_3)
\end{equation}
This orbifold is singular on the $x_1=0$ plane, has a quadrant-size fundamental domain and can be used for head-on collision and single holes but not for rotating binary systems.

\section{The {\it Octant} orbifold}

In the octant orbifold configuration, the group of symmetries is generated by the maximum number of three space reflections. Without loss of generality, this group is 

\begin{equation}
G_o=\langle r_{x_1}, r_{x_2}, r_{x_3} \rangle
\end{equation}
where $r_{x_i}$ is the reflection
\begin{equation}
x_i=-x_i
\end{equation}

This orbifold will have singularities on the $x_1=0$, $x_2=0$ and $x_3=0$ planes and their intersections. The fundamental domain of the orbifold is the octant $x_1>0, x_2>0, x_3>0$. {\it Octant} has been used for evolving single black holes, black holes perturbed by Brill Waves \cite{Garfinkle:2000hd}, non-rotating equal mass black holes colliding head-on  and many other configurations.  The octant orbifold provides a cheap and very useful testbed for numerical simulations. Results obtained in Octant symmetry are mentioned in \cite{Alcubierre:2002kk, Laguna:2002zc, Yo:2002bm, Alcubierre:2000yz, Alcubierre:1999ab, New:2000dz, Bona:1998dp, Alcubierre:2001vm, Brugmann:1996kz, Bondarescu:2001jf}

\subsection{Alternatives to {\it Octant}}

An alternative orbifold that can be used in many applications and has the same fundamental domain as the {\it Octant} orbifold is given by the group of isometries 

\begin{equation}
G=\langle r_{x_3}, (x_1, x_2, x_3) = (-x_2, x_1, x_3)  \rangle
\end{equation}

Unlike {\it Octant}, this orbifold permits rotation in the $(x_1,x_2)$ plane and due to its different set of symmetries it will allow a different set of interesting/unstable modes to grow.

\section{Periodic boundary Condition as an orbifold}

Periodic boundary conditions are another usual setting for Numerical Relativity simulations. To obtain periodic boundary condition, one identifies 
\begin{eqnarray}
x_1=x_1+L_{x_1} \nonumber \\
x_2=x_2+L_{x_2}
\label{periodicBC} \\
x_3=x_3+L_{x_3}  \nonumber
\end{eqnarray}
The identifications (\ref{periodicBC}) generate an infinite group isomorphic to ${\bf Z}\times {\bf Z} \times {\bf Z}$ . Moding $R^3$ by this group yields $T^3$, the three dimensional torus, an orbifold often used in simulations \cite{Shinkai:2000eu, Bona:2002yy, Calabrese:2004qx, Alcubierre:2003pc, Szilagyi:1999nu}. The main advantage of $T^3$ is that is has no boundary and finding good boundary conditions is still an open problem in Numerical Relativity.

The main shortcoming of $T^3$ is that usually the space-time it describes is not asymptotically flat.

\section{Dimensional Reduction via orbifolding}
 
The group $G$ in the definition of the orbifold (\ref{coset}) does not have to be finite. It can even be a manifold itself. For example, if one chooses $G$ to be the group of rotations $SO(3)$ and $M$ the 4 dimensional Minkowski space, the orbifold $O=M/G$ will only be two-dimensional. If one takes time out of the picture and considers $M$ to ${\bf R}^3$, the orbifold ${\bf R}^3/SO(3)$ will only have one dimension. This orbifold is often used for evolving single non-rotating black holes and provides a valuable cheap tool for code testing. 

Orbifolding via the group of rotations around a given axis $SO(2)$ gives a $2+1$ dimensional arena where axisymmetric black hole collisions can be studied. Valuable results of simulations carried out on this orbifold have been published in 
\cite{Anninos:1994pa}, \cite{Anninos:1994ay}, \cite{Masso:1998fi}.

Some space-times, like orbiting black holes or neutron stars, have approximate helical symmetry. Forcing this symmetry to be exact, and moding the space-time by the 1-dimensional group generated by this symmetry yields ${\bf R}^3$ as an orbifold of Minkowski. Thus, time is eliminated and the 4D problem is reduced to a 3D one. The method is called "quasi-equilibrium approximation" and has been reported in \cite{Beetle:2006rd, Gourgoulhon:2000nn, Gourgoulhon:2001ec, Klein:2004be, Friedman:2001pf}.

\section{Summary}

In this paper, we discussed orbifolds that have been used in Numerical Relativity and orbifolds that could be used, but have not been used yet. In the past many authors discussed simulations of the same physical problem on several orbifolds with different fundamental domanins like {\it octant, bitant, quadrant} and full ${\bf R^3}$.  Running simmulations for the same physical problem on several different orbifolds with the same fundamental domain will allow Numerical Relativists to gain more insight into general relativity and provide useful testbeds for simmulations. 

On the analytical front, orbifolds provide a plethora of new exact solutions of the Einstein Equations. Most of them can be trivially obtained by restricting metrics with a given symmetry $s$ to the orbifold $O=M^4/G$ where $G=\langle s \rangle$ is the group generated by $s$. In the case of reflection symmetry, we obtain {\it Bitant}.  One exact solution would be a Schwarzschild metric centered at the origin of {\it Bitant}. This metric trivially satisfies The Einstein Equations, but the solution, being an orbifold, has a different topology then the regular full Schwarzschild 
solution it was derived from. 

\section{Acknowledgments}

The author thanks Gabrielle Allen, Miguel Alcubierre, Gregory Daues, Nathan Dunfield,  Sergei Gukov, Graeme Smith and Ed Seidel for useful discussions, advice support and friendship. 

Eric Gourgoulhon and Erik Schnetter  have provided valuable comments on the first version of the manuscript.  

This work was supported by a Caltech Teaching Assistantship with David Politzer.

\end{document}